**Model of multilayered materials for interface stresses estimation**

**and validation by finite element calculations**

**R. P. Carreira, J. F. Caron*, A. Diaz Diaz**

Laboratoire d'Analyse des Matériaux et Identification

Ecole Nationale des Ponts et Chaussées

http://www.enpc.fr  *caron@lami.enpc.fr

6 et 8 avenue Blaise Pascal – Cité Descartes – Champs-sur-Marne

77455 Marne-La-Vallée Cedex 2, France

**Abstract**

*The mechanical problem discussed in this paper focuses on the stress state estimation in a composite laminate in the vicinity of a free edge or microcracks. To calculate these stresses, we use two models called Multiparticle Models of Multilayered Materials (M4). The first one can be considered as a stacking sequence of Reissner-Mindlin plates (5 kinematic fields per layer), while the second is a membranar superposition (2 fields per layer plus a global one). These simplified models are able to provide finite values of interfacial stresses, even on the free edges of a structure. The current paper consists of validating the M4 by a finite element analysis through describing the stress fields in both a $(0,90)_s$ laminate in tension (free-edge problem) and a transversally microcracked $(0,90)_s$ laminate. A comparison of the various energy contributions helps yield a mechanical perspective: it appears possible to define an interply energy as well as a layer energy, these energies expressing the FE 3D reality.*

## 1. Introduction

Due to the large difference in anisotropy of two consecutive plies, high interlaminar stresses (at the interplies and especially in the vicinity of a free edge) are induced in cross-ply laminates and lead to damage such as delamination. Classical lamination theory is not able to calculate these out-of-plane stresses. Many studies have sought to overcome this lack of classical lamination theory by calculating the interlaminar stresses in a laminate subjected to a tensile loading. A bibliography reviewing each of these studies in detail is detailed in this paper. They include numerical and analytical studies, the finite difference technique, the finite element method, boundary layer theories with corrective terms and models with a kinematic field per layer, that we named multiparticle models.

The models proposed herein, called Multiparticle Models of Multilayered Materials (M4), clearly belong to this last family of models. Their construction will be summarily described.

In order to validate our models, two examples corresponding to different boundary conditions are treated. The first one consists of analysing the free edge problem in a $(0,90)_s$ laminate submitted to tension. The second considers, in the same laminate, microcracks present within the 90° plies. These two examples, as well as all materials constants and calculation hypotheses, are depicted in the following section.. The use of such cross-ply allows us to avoid 3D calculations by admitting plane deformation assumptions, and hence 2D calculations. The study of a $(\theta,-\theta)_s$ will enable drawing the exact same kind of observations and conclusions, but with 3D meshing (see Figure 12, for example).

Once the validity of the finite element calculations (numerical convergence and study of singularities) has been ensured, a comparison of the two approaches is drawn in the paper's final section, focusing on:: 3D stress fields, interface forces, and proposing energy-related considerations. In the case of free edge problem, our results also make reference to some of Pagano's works (1978).

## 2. Stress calculation methods

In this section, we will present some of the methods available for calculating interlaminar stresses in any laminate with straight free edges. These consist of: finite element procedures, boundary layer theories and multiparticle models.

Wang & Crossman (1977) used a finite element numerical procedure based on a displacement formulation. The field singularities between two plies and near the free edge are highlighted. For the crossplies ($(0/90)_s$, $(90/0)_s$), they established a description (on the mid-plane) of $\sigma_{zz}$ that displays a different sign between the two cases. These various curves for $\sigma_{zz}$ may serve to justify the distinction in the two laminates' behaviour regarding damage at this interface. At the 0°/90° interface, shear stress conditions are also different: for the $(0/90)_s$ laminate, a singularity seems to exist at the free edge, which is not the case for the $(90/0)_s$ laminate. Raju & Crews (1981) investigated the $(\theta,\theta-90)_s$ family. With a refined polar mesh, they were able to determine the stress singularity order. This singularity was studied by a number of authors (Wang & Choi (1982a) , Leguillon (1998)), etc.) and has often been identified using a logarithmic expression. Shah & Murty (1991) modelled the laminate as a combination of three distinct regions: quasi-3D elements close to the free edge, linear elements over the entire plate, and transitional elements between these two regions. Robbins & Reddy (1993) developed a 2D layerwise, displacement-based finite element model of laminated composites that assumes a per-layer distribution of the displacement field (1D elements on the thickness).

Because classical lamination models yield an accurate approximation of fields except in the vicinity of edges, it appears altogether natural to superimpose corrective fields whose values are only significant at the edges (boundary layer theory) onto these lamination fields. Let us first point out the asymptotic development technique by Lécuyer (1991) and the Fourier series development of Allix (1992). Wang & Choi (1982b) presented a formulation of stress functions based on complex variables. The boundary layer asymptotic stresses are characterised by introducing a stress intensity factor dependent upon geometrical variables (e.g. laminate thickness, number of plies), stacking parameters (fibre orientation, stacking sequence) and environmental conditions (temperature and relative humidity).

Our attention is now turned to defining a family of models we call multiparticle models; these consider the existence of many material particles at a single geometrical point, i.e. one particle (or one kinematic) per layer (whereas classical lamination theory involves only one kinematic over the entire thickness). Nevertheless, they are all 2D models and, as such, can be viewed as plates or membranar superpositions linked together by interface forces. Garett & Bailey (1977) developed a model that enables solving transverse-cracking problems in a $(0/90)_s$ laminate. This model, called shear-lag, has been widely used by other authors (Caron et Ehrlacher (1997,1998), Carreira(1998), Steif (1983). The most complete multiparticle model is the local model of Pagano (1978). He proposed a laminated composite theory based on the Hellinger-Reissner variational principle (1950): the membranar stresses in each ply are written as first-order polynomials, and the shear and normal stresses are then obtained by integrating the 3D equilibrium equations. The various studies of Caron & Ehrlacher (1997), Chabot (1997), Naciri et al. (1998), 1998), Caron et al. (1999), Carreira (1998), (referring to the **M4**) take their inspiration from Pagano's model (1978). However they are aimed at proposing more simplified approaches, which serve to derive analytical solutions.

**3. Presentation of the two problems : free edge and micro-cracking boundary conditions**

Each ply is considered as macroscopically homogeneous and monoclinic, and represented by its elastic constants : $E_l$, $E_t$, $E_n$ longitudinal, transverse and normal Young's moduli, $G_{ij}$ shear moduli and $\nu_{ij}$ Poisson coefficients.

The first example consists of the free edge problem in a finite width $(0,90)_s$ laminate under uniaxial tension (see *figure 1*), where e denotes the ply thickness, and 2*b* the width. Using the relation *b*=8*e* as a base asumption, a uniform displacement ±Δ is then imposed on *x*=±*a* edges. For the stacking sequence studied herein, the problem is independent of *x* and due to symmetry, the problem can be reduced to an analysis of the shaded cell in the (*y*,*z*) plane.

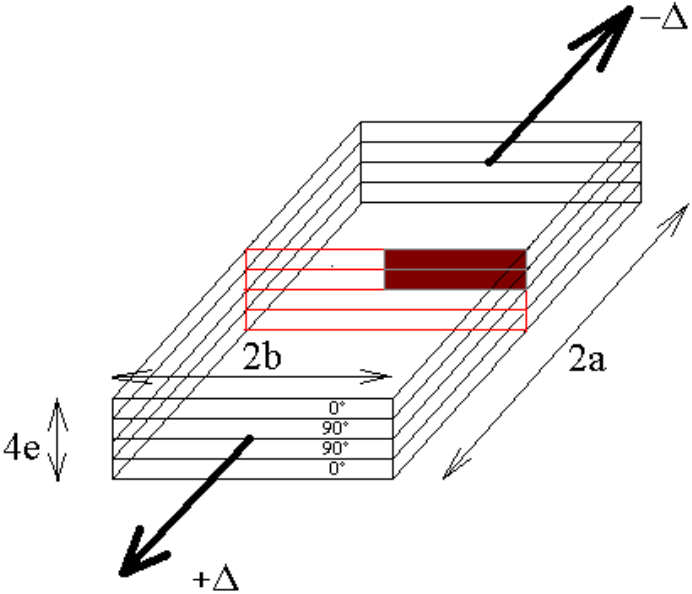

*fig 1 : The free edge problem in a $(0,90)_s$ laminate under uniaxial tension*

The second example studied, which depicts more severe stress gradients conditions, is transverse cracking in a $(0,90)_s$ laminate under uniaxial tension in the *x*-direction (see Figure 2). A periodic cell (in the (*x*,*z*) plane) is highly representative of this problem and a mean distance (2*h*) between two consecutive cracks can in reality be experimentally observed. In comparison with the free edge problem, only boundary conditions have changed: instead of an out-of-plane loading, this second problem now consists of a prescribed displacement of the 0° ply, with the 90° ply remaining free of stresses. Note that we have selected $b = h = 8e$ in order to ensure cells of the same dimensions.

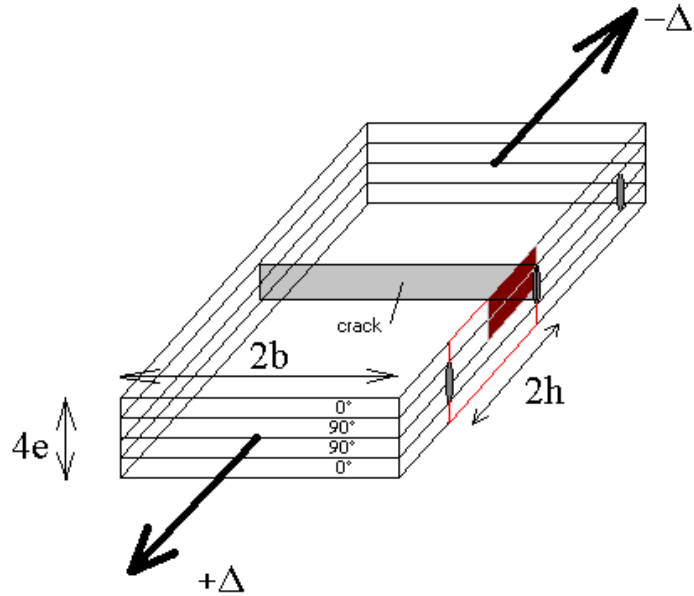

*fig 2 : The problem of a (0,90)ₛ laminate under uniaxial tension with 90° ply micro-cracks*

The material properties and sample geometry are summarised in Table 1 below:

| **Material : carbon-epoxy** | **Dimensions** |
|---|---|
| $E_l = 137.9 \ GPa, \ E_t = E_n = 14.48 \ GPa$ | $e = 0,14 \ mm$ |
| $G_{lt} = G_{ln} = G_{nt} = 5.86 \ GPa$ | $b = h = 8e$ |
| $\nu_{lt} = \nu_{ln} = \nu_{nt} = 0.21$ | $a \approx 20b$ |

*Table 1 : material properties (Wang & Crossman (1977)) and sample dimensions*

## 4. The M4 construction

We introduce the following notations : $x$ and $y$ represent the co-ordinates in the mid-plane of the layer, $z$ is the thickness co-ordinate. In each layer $i$ ($i$=1,$n$), $h_i^-$, $h_i^+$ and $\bar{h}_i$ are the bottom, the top and the mid-plane $z$ co-ordinates of the ply, respectively, and $e_i = h_i^+ - h_i^-$ is the thickness. Greek alphabet subscripts correspond to {1,2} and Latin to {1,3} (except for $i$ which identifies the layer).

The M4 construction method (Chabot (1997)) is based upon the four steps described below. The M4_2$n$+1 is the simplest one : it considers the laminate as a membranar superposition (2$n$ equations plus a global one, with $n$ being the number of plies in the laminate). Resultant forces in each layer as well as interlaminar shear stresses are taken into account, yet resultant moment in the layer is not. The M4_5n can be considered as a superposition of Reissner-Mindlin plates, in taking the shear and moment resultant, in each layer, into account, along with interlaminar shear and normal stresses at the interfaces ($5n$ equilibrium equations).

## 4.1 Three-dimensional stress approximation

The first step consists of writing an approximation of the 3D stress fields as $z$-dependent polynomials within each layer. The polynomial coefficients are functions of $x$ and $y$ only and are expressed in terms of what we call generalised internal forces (defined in each layer or at the interfaces) and their number govern the wealth (but also the complexity) of the final model. In this way, one of our models (M4_7n) for example, corresponds to Pagano's model (1978). The name of each model is derived, as stated above, from the number of equilibrium equations to be satisfied.

### *The M4_5n model*

The in-plane stress components $\sigma_{\alpha\beta}$ ($\alpha,\beta \in \{1,2\}$) are chosen as linear functions of $z$ and the 3D equilibrium equations lead both to shear stresses $\sigma_{\alpha3}$ in the form of quadratic polynomials of $z$ and to the normal stress $\sigma_{33}$ as third-order polynomials. The polynomial coefficients are expressed in terms of the following generalised internal forces :

- force, moment and shear resultants tensors of layer $i$, respectively :

$$N_{\alpha\beta}^i(x,y) = \int_{h_i^-}^{h_i^+} \sigma_{\alpha\beta}(x,y,z)\,dz \qquad ; \qquad M_{\alpha\beta}^i(x,y) = \int_{h_i^-}^{h_i^+} (z - \overline{h}_i)\sigma_{\alpha\beta}(x,y,z)\,dz$$

$$Q_\alpha^i(x,y) = \int_{h_i^-}^{h_i^+} \sigma_{\alpha3}(x,y,z)\,dz \qquad\qquad (1)$$

- interlaminar shear and normal stresses at interfaces $i,i+1$ and $i-1,i$ :

$$\tau_\alpha^{i,i+1}(x,y) = \sigma_{\alpha3}(x,y,h_i^+) \qquad\qquad \tau_\alpha^{i-1,i}(x,y) = \sigma_{\alpha3}(x,y,h_i^-) \qquad\qquad (2)$$

$$\nu^{i,i+1}(x,y) = \sigma_{33}(x,y,h_i^+) \qquad\qquad \nu^{i-1,i}(x,y) = \sigma_{33}(x,y,h_i^-) \qquad\qquad (3)$$

### *The M4_2n+1 model*

The previous model may be simplified both by neglecting the moment resultants in each layer $i$ (very thin layers) and by excluding interlaminar normal stresses (eq. 3): the $\sigma_{\alpha\beta}$ stresses are therefore independent of the $z$ co-ordinate. The approximation order of $\sigma_{\alpha3}$ and $\sigma_{33}$ must remain consistent with the equilibrium equations (first and second order polynomials, respectively).

The M4_5$n$ and M4_2$n$+1 generalised internal forces are summarised in Figures *3*.

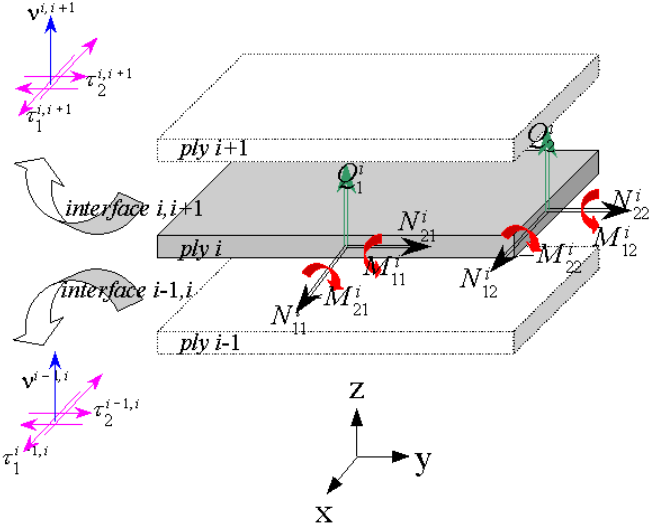

*Fig 3a : M4_5n generalised internal forces*

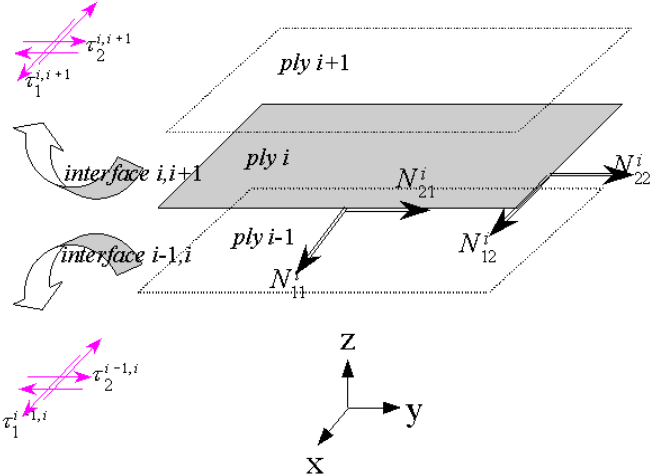

*Fig 3b : M4_2n+1 generalised internal forces*

## 4.2 Associated generalised displacements and deformations

The assumed stress fields are inputted into the following Hellinger-Reissner functional (H.R.F.) , Reissner (1950)

$$H.R.\left(\underline{U}^*,\underline{\underline{\sigma}}^*\right)=\int_{\Omega}\left[\underline{\underline{\sigma}}^*:\underline{\underline{\varepsilon}}\left(\underline{U}^*\right)-\underline{f}.\underline{U}^*-\frac{1}{2}\underline{\underline{\sigma}}^*:\underline{\underline{\underline{S}}}:\underline{\underline{\sigma}}^*\right]d\Omega-\int_{\partial\Omega_u}\left(\underline{\underline{\sigma}}^*.\underline{n}\right).\left(\underline{U}^*-\underline{U}^d\right)dS-\int_{\partial\Omega_t}\underline{T}^d.\underline{U}^*dS$$

Where the displacement $\underline{U}^*$ is a field of a continuous vector, whereas the 3D stress $\underline{\underline{\sigma}}^*$ is a field of a symmetrical second-order tensor. $\Omega$ is the studied object volume, $\partial\Omega$ its boundary, $\underline{\underline{\varepsilon}}\left(\underline{U}^*\right)$ is the symmetrical gradient of $\underline{U}^*$, $\underline{\underline{\underline{S}}}$ is the compliance tensor, $\underline{T}^d$ is a surfacic force on $\partial\Omega_t$ (part of $\partial\Omega$), $\underline{U}^d$ is a prescribed displacement on $\partial\Omega_u$ (part of $\partial\Omega$), and $\underline{n}$ is the normal to $\partial\Omega$.

After integration in each ply with respect to $z$, these associated generalised displacements are then deduced (see Chabot (1997) for more details). They appear as weighted-average 3D displacements. For the M4_5n, we can define the following 5n fields :

- the in-plane displacement and rotation fields of layer $i$, whose components are :

$$U_\alpha^i\left(x,y\right)=\frac{1}{e^i}\int_{h_i^-}^{h_i^+}U_\alpha\left(x,y,z\right)dz\qquad;\qquad\Phi_\alpha^i\left(x,y\right)=\frac{12}{e^{i^2}}\int_{h_i^-}^{h_i^+}\frac{z-\overline{h_i}}{e^i}U_\alpha\left(x,y,z\right)dz\qquad(4)$$

- and the vertical displacement of layer $i$, $U_3^i$ such that:

$$U_3^i\left(x,y\right)=\frac{1}{e^i}\int_{h_i^-}^{h_i^+}U_3\left(x,y,z\right)dz\qquad(5)$$

By noting $W_3=\dfrac{U_3\left(h_n^+\right)+U_3\left(h_1^-\right)}{2}$ the 2n+1 generalised displacements of the M4_2n+1 are identified as $U_\alpha^i$ and $W_3$.

The M4_5n and M4_2n+1 generalised displacements are summarised in Figure 4.

The generalised strains, deduced from the generalised displacements, appear as the cofactors of the generalised internal forces in the Hellinger-Reissner functional. For the M4_5n, $N_{\alpha\beta}^i$, $M_{\alpha\beta}^i$, $Q_\alpha^i$, $\tau_\alpha^{i,i+1}$ and $\nu^{i,i+1}$ are associated with $\varepsilon_{\alpha\beta}^i$, $\chi_{\alpha\beta}^i$, $d_{\Phi\alpha}^{\;i}$, $D_\alpha^{i,i+1}$ and $D_3^{i,i+1}$ respectively which, are defined as follows :

$$\varepsilon_{\alpha\beta}^i=\frac{1}{2}\left(U_{\alpha,\beta}^i+U_{\beta,\alpha}^i\right)\quad;\quad\chi_{\alpha\beta}^i=\frac{1}{2}\left(\Phi_{\alpha,\beta}^i+\Phi_{\beta,\alpha}^i\right)\quad;\qquad d_{\Phi\alpha}^{\;i}=\Phi_\alpha^i+U_{3,\alpha}^i$$

$$D_\alpha^{i,i+1}=U_\alpha^{i,i+1}-U_\alpha^i-\frac{e^i}{2}\Phi_\alpha^i-\frac{e^{i+1}}{2}\Phi_\alpha^{i+1}\qquad;\qquad D_3^{i,i+1}=U_3^{i+1}-U_3^i\qquad(6)$$

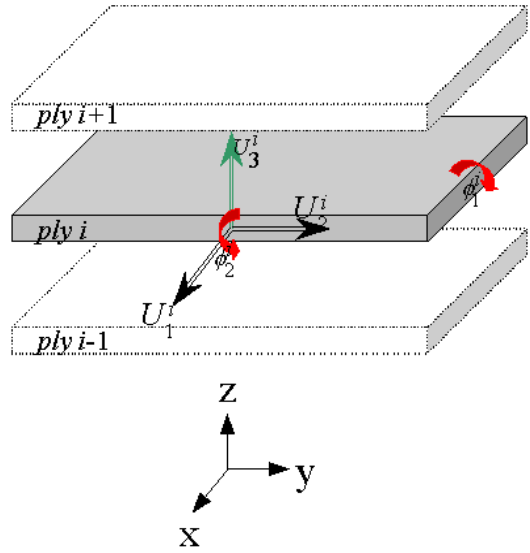

*Fig 4a : M4_5n generalised displacements*

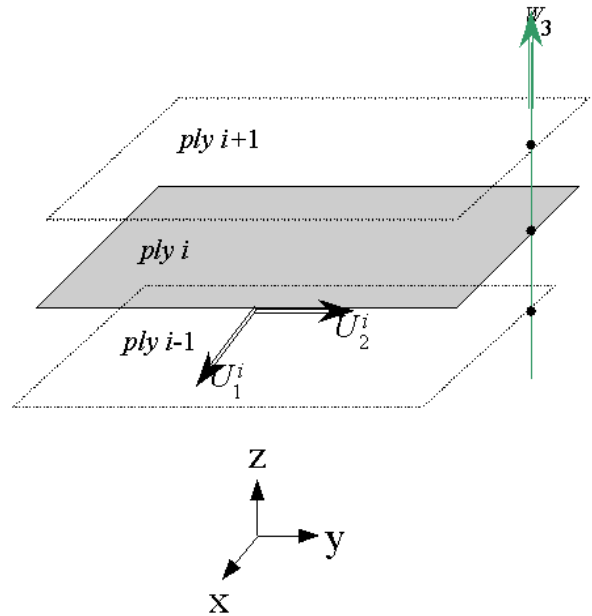

*Fig 4b : M4_2n+1 generalised displacements*

For the M4_2n+1, $N_{\alpha\beta}^i$ and $\tau_\alpha^{i,i+1}$ remain associated with $\varepsilon_{\alpha\beta}^i$ and $D_\alpha^{i,i+1}$ respectively, whereby:

$$D_\alpha^{i,i+1} = U_\alpha^{i,i+1} - U_\alpha^i + \frac{e^i + e^{i+1}}{2} W_{3,\alpha} \qquad (7)$$

## 4.3 Equilibrium equations

*Reissner (1950): The elastic solution of the problem is the pair $\left(\underline{U}^*,\underline{\underline{\sigma}}^*\right)$which renders the Hellinger-Reissner functional stationary.*

Hence, the derivation of the functional with respect to generalised displacement fields leads to the equilibrium equations of each of the approximate models, which in turn leads respectively to the following $5n$ and $2n+1$ equilibrium relations ($\alpha,\beta \in \{1,2\}$) :

$$5n \begin{cases} N^i_{\alpha\beta,\beta} + \left(\tau^{i,i+1}_{\alpha} - \tau^{i-1,i}_{\alpha}\right) = 0 \\ Q^i_{\beta,\beta} + \left(\nu^{i,i+1} - \nu^{i-1,i}\right) = 0 \\ M^i_{\alpha\beta,\beta} + \dfrac{e^i}{2}\left(\tau^{i,i+1}_{\alpha} + \tau^{i-1,i}_{\alpha}\right) - Q^i_{\alpha} = 0 \end{cases} \qquad 2n+1 \begin{cases} N^i_{\alpha\beta,\beta} + \left(\tau^{i,i+1}_{\alpha} - \tau^{i-1,i}_{\alpha}\right) = 0 \\ \displaystyle\sum_{j=1}^{n}\left(\dfrac{e^j}{2}\left(\tau^{j,j+1}_{\beta,\beta} + \tau^{j-1,j}_{\beta,\beta}\right)\right) + \left(\nu^{n,n+1} - \nu^{0,1}\right) = 0 \end{cases} \quad (8)$$

## 4.4 Constitutive equations

The derivation of the Hellinger-Reissner functional with respect to the generalised force fields yields each model's constitutive equations. The ply is assumed monoclinic and $S_{mnop}$ represents the components of the compliance tensor with adapted symmetries.

Thus, we can deduce the generalised strains and forces relationship as follows:

- bending and torsion behaviour of layer $i$ :

$$\varepsilon^i_{\alpha\beta} = \frac{1}{e^i}S^i_{\alpha\beta\gamma\delta}N^i_{\gamma\delta} \qquad ; \qquad \chi^i_{\alpha\beta} = \frac{12}{e^{i3}}S^i_{\alpha\beta\gamma\delta}M^i_{\gamma\delta} \qquad (9)$$

- the out-of-plane shear behaviour of layer $i$ :

$$d_{\Phi\alpha}{}^i = f\left(Q^i_{\beta}, \tau^{i,i+1}_{\beta}, \tau^{i-1,i}_{\beta}, S^i_{\alpha3\beta3}\right) \qquad (10)$$

- the behaviour of the interlaminar shear and normal stresses at interface $i,i+1$ :

$$D^{i,i+1}_{\alpha} = f\left(Q^i_{\beta}, Q^{i+1}_{\beta}, \tau^{i-1,i}_{\beta}, \tau^{i+1,i+2}_{\beta}, \tau^{i,i+1}_{\beta}, S^i_{\alpha3\beta3}, S^{i+1}_{\alpha3\beta3}\right) \qquad (11)$$

$$D^{i,i+1}_{3} = f\left(\nu^{i-1,i}, \nu^{i,i+1}, \nu^{i+1,i+2}, S^i_{3333}, S^{i+1}_{3333}\right) \qquad (12)$$

Lastly, we can write for the M4_2$n$+1 :

$$\varepsilon^i_{\alpha\beta} = \frac{1}{e^i}S^i_{\alpha\beta\gamma\delta}N^i_{\gamma\delta} \qquad ; \qquad D^{i,i+1}_{\alpha} = f\left(\tau^{i-1,i}_{\beta}, \tau^{i+1,i+2}_{\beta}, \tau^{i,i+1}_{\beta}, S^i_{\alpha3\beta3}, S^{i+1}_{\alpha3\beta3}\right) \qquad (13)$$

The complete expressions for $d_{\Phi\alpha}{}^i$, $D^{i,i+1}_{\alpha}$ and $D^{i,i+1}_{3}$ can be found in Chabot (1997).

## 5. Numerical aspects

This section is intended to validate our finite element results, which are our main reference during the steps of M4 validations, and serves to introduce what we have called the *finite element generalised interface forces*.

## 5.1 Numerical convergence and singularities

As a means of displaying the numerical convergence of our finite elements calculations, we have examined study shear stresses in the case of the free edge problem. Our finite element analysis merely allows identifying stresses values in each element and not, unfortunately, those located exactly at the interface, as do our simplified models (see equations 2 and 3). It is therefore necessary to calculate the mean stresses ($\frac{\sigma_{abo} + \sigma_{bel}}{2}$) at the interface by taking values just *above* and just *below*. Shear stress curves have been plotted for three different meshes (corresponding to the shaded cells in Figures 1 and 2) : **50x12**, **25x6** and **16x4** meshes, respectively (see Figure 5).

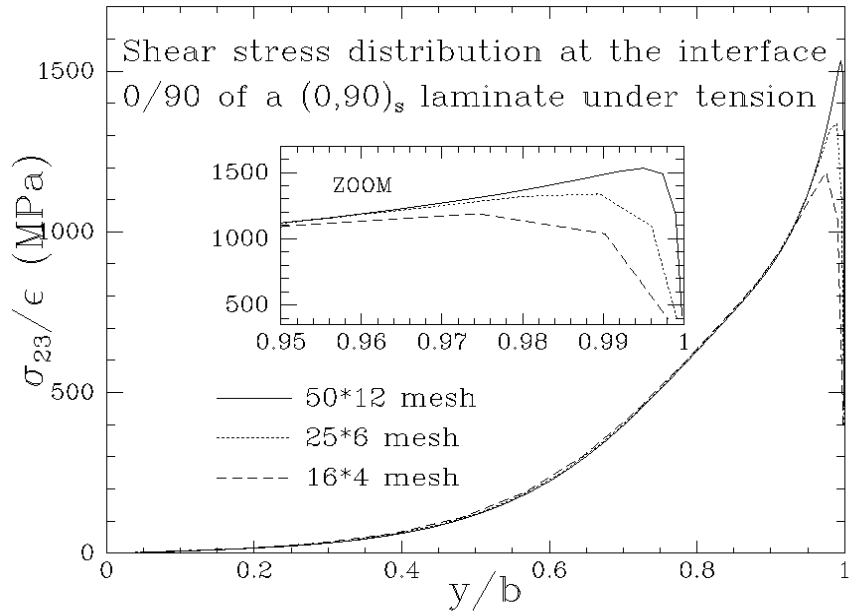

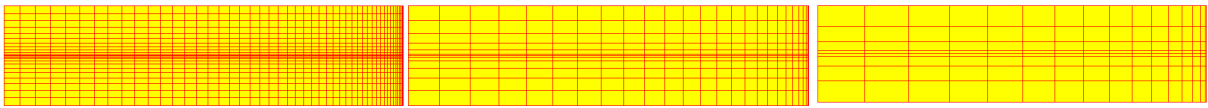

*Fig 5 : Mesh influence on the finite element mean shear stress*

For the first mesh for example, this reflects **12** elements per ply thickness and **50** elements in the width direction. According to figure *5*, shear stress singularity at the free edge exists. In effect, as the mesh becomes finer, distance to the edge decreases and shear values increase. As a case in point, the value for a **50x12** mesh is about 35% greater than that for a **25x6** mesh. This maximum value is thereby rendered meaningless due to mesh dependence.

Because of the steepness of stress gradients at the ply interface, particularly near the free edge (see Figure 6), it can be observed that the coarser mesh only shows convergence (i.e. discrepancy of less than 1%) for stress values (*above* and *below*) up to $y/b$=0.92. In refining the mesh (*i.e.* the **50x12** mesh), convergence only occurs at a distance of about 2% (between $y/b$=0.98 to $y/b$=1) of the cell.

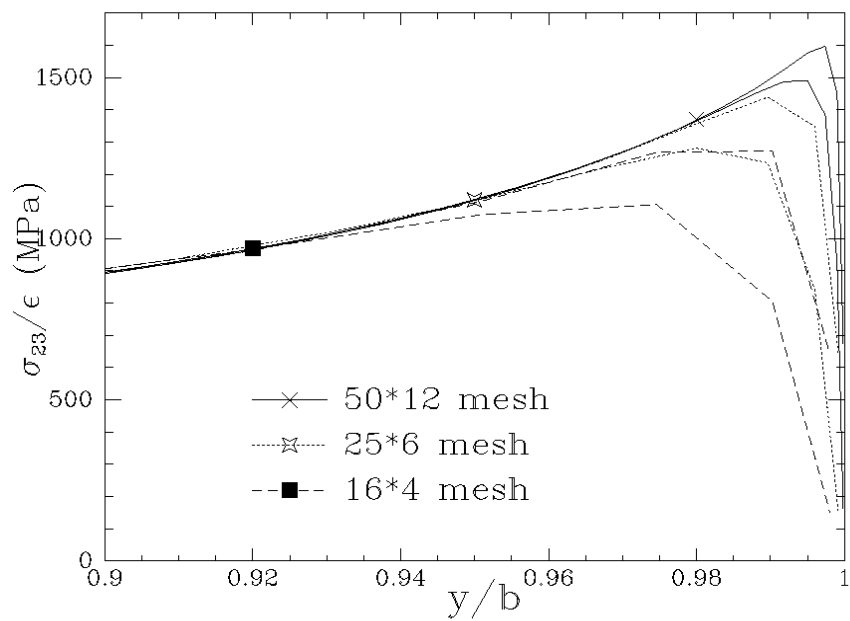

*Fig 6 : Convergence between above and below FE shear stresses*

Two main difficulties have arisen: 1)identifying the stresses near the edge, and 2)calculating those located exactly at the interface. In order both to overcome these difficulties and to draw a comparison with our simplified models, we have introduced what we call the *finite element generalised interface forces*, which are very similar to those introduced for the M4 (see equations 1 through 3).

## 5.2 Definition of the finite element generalised internal forces

We are introducing here this particular force concept in order to determine the finite element forces specially at the interfaces and to better describe these forces near the edge. Let's write out the first two equilibrium equations of M4_5n (equation 8 left) :

$$N_{\alpha\beta,\beta}^i + \left(\tau_\alpha^{i,i+1} - \tau_\alpha^{i-1,i}\right) = 0 \qquad ; \qquad Q_{\beta,\beta}^i + \left(\nu^{i,i+1} - \nu^{i-1,i}\right) = 0$$

By summing over the first $j$ plies, we obtain the following expressions for the interlaminar shear and normal stresses at interface $j,j+1$ (no surfacic force has been prescribed) :

$$\tau_\alpha^{j,j+1} = -\sum_{i=1}^{j} N_{\alpha\beta,\beta}^i \qquad ; \qquad \nu^{j,j+1} = -\sum_{i=1}^{j} Q_{\beta,\beta}^i \qquad (14)$$

For a $(0,90)_s$ laminate submitted to uniaxial tension, the finite element shear and normal stresses ($\tau_2^{0,90\,FE}$, $\nu_{0,90}^{FE}$ and $\nu_{90,90}^{FE}$, respectively), can be deduced by deriving $N_{22}^i$ and $Q_2^i$ numerically :

$$\tau_2^{0,90\,FE} = \sigma_{23}(x,y,z=e) = -\frac{\partial N_{22}^{90}(y)}{\partial y} \quad ; \qquad \nu_{0,90}^{FE} = \sigma_{33}(x,y,z=e) = \frac{\partial Q_2^0}{\partial y}$$

$$\nu_{90,90}^{FE} = \sigma_{33}(x,y,z=2e) = \frac{\partial Q_2^0}{\partial y} + \frac{\partial Q_2^{90}}{\partial y} \qquad (15)$$

$\tau_\alpha^{FE}$ and $\nu^{FE}$ are referred to as the finite element generalised internal forces.

As undertaken previously for the mean shear stresses, we now conduct a convergence study for the generalised shear stresses. The curves in Figure 7 attest to the convergence of $\tau_2^{FE}$, regardless of the mesh used, as long as $\frac{y}{b} < 0.98$.

The finite element generalised internal forces we defined are in fact more effective and pertinent in estimating 3D stresses near the edge.

### 5.3 Definition of a zone of confidence

It is appropriate, for the example treated herein, to compare this convergence distance (2% of the cell away from the edge) with the dimension of material's constitutive carbon fibres (cf. figure 8).

In recalling that the fibre diameter $d_f$ is equal to approximately 7 μm, the difference noted between shear stresses only affects a distance of about $3d_f$. Thus, it doesn't seem highly useful to focus on elements so close to the edge. For such distances, fibre and resin behaviour should be studied separately (Kassapoglou & Lagace (1986)) and the hypothesis of material being macroscopically homogeneous is no longer valid.

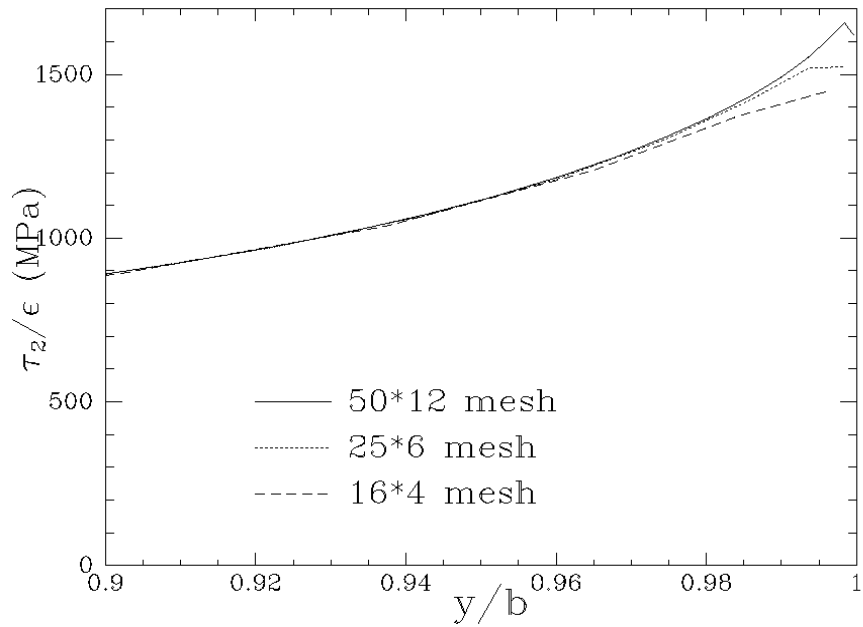

*Fig 7 : Convergence of the finite element generalised shear stresses*

For the present case, we can apply the coarsest mesh that still yields a convergent result at a distance of up to 20 μ*m* from the edge.

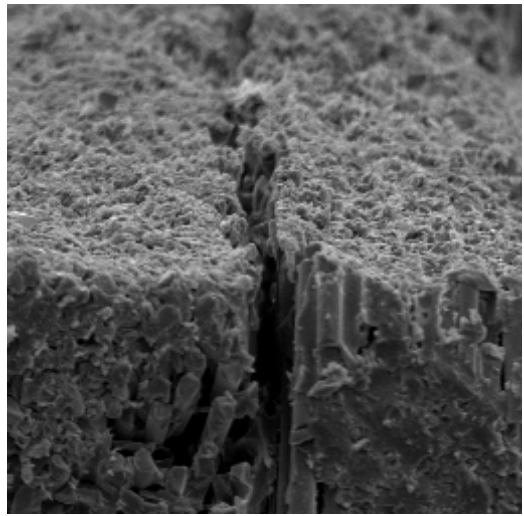

*Fig 8 : MEB photography of an edge of CFRP laminate*

Let us recall that our simplified models produce finite values for interface forces, even on the edge; however for purposes of consistency, we will, throughout the following, always be comparing FE results and M4 analytical solutions only over what has been designated a *zone* *of confidence* (i.e. where finite element calculations are not mesh-dependent. Only over such a zone therefore are our models able to be validated. Nevertheless, the pertinence of finite values on the edges of M4 interface forces may be improved by complementary approach, e.g. experimentation (see Caron *et al.* (1999)).

As a conclusion to this section on numerical aspects, the interest of finite element generalised internal forces has been clearly demonstrated: the stresses are calculated at exactly the interface and convergence of the results is better ensured.

## 6. The M4 validation : three steps for validating models

In this section, our goal is to compare, over the zone of confidence, finite element results with M4 analytical solutions through, in particular, a 3D stress fields comparison (throughout the laminate thickness), and then the interface stresses (finite element generalised forces). Finally, a study on energy distributions is conducted. It should be pointed out that the M4_$5n$ analytical solution to the micro-cracked $(0,90)_s$ laminate can be found in Carreira (1998).

### 6.1 Three-dimensional fields

Our primary aim here is to plot the shear stresses with respect to laminate thickness for several distances from the edge (see Figure 9a for the free edge, Figure 9b for the microcrack), in order to compare the three solutions (finite element, M4_$5n$ and M4_$2n$+1). This approach thereby allows measuring the error introduced in choosing an approximated $z$-polynomial field to describe 3D reality (linear for the M4_$2n$+1 and second order, for the M4_$5n$).

The essential findings are as follows:

- A very strong correlation has been noted between finite element and M4_$5n$ distributions except in the vicinity of the micro-crack, due to the steep stress gradient.

- The approximated shear stresses are linearly dependent upon $z$ and symmetrical throughout the interface for the M4_$2n$+1 : consequently, the correlation is not very strong.

- The results coincide perfectly at the interface, for both models, with the exception of M4_$2n$+1 in the case of the free edge problem.

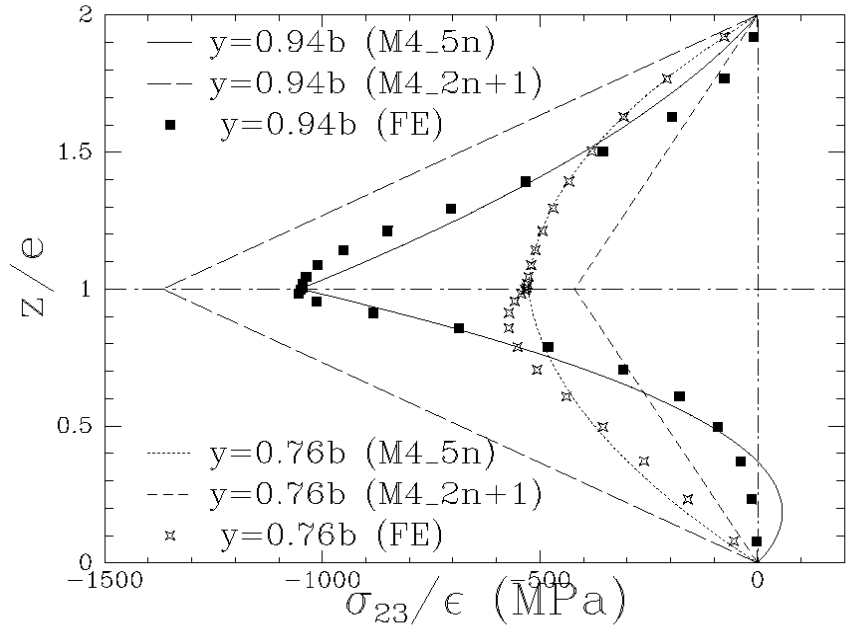

*Fig 9a : Shear stress distributions as a function of z/e for different distances from the edge.*

*Free edge problem*

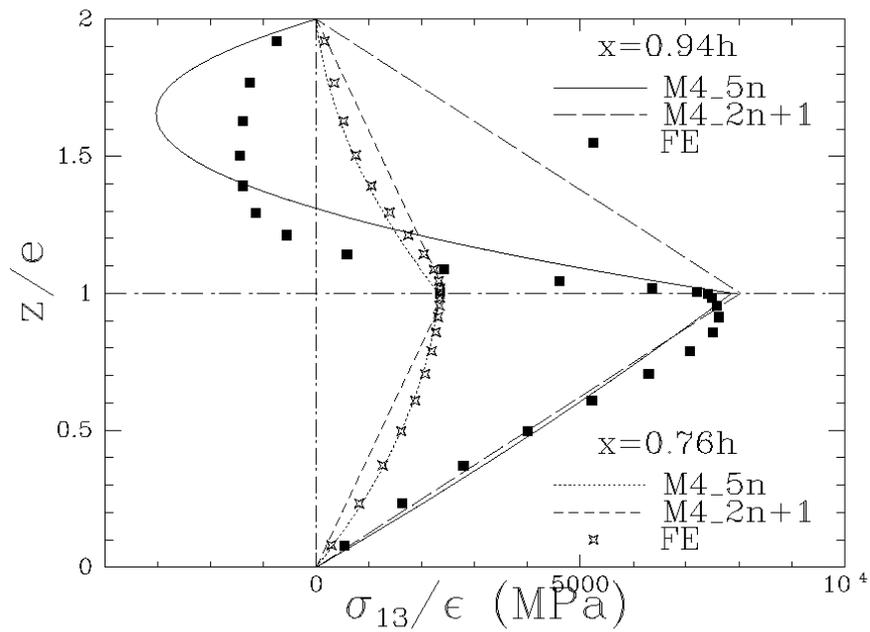

*Fig 9b : Shear stress distributions as a function of z/e for different distances from the edge.*

*Micro-cracked cell*

## 6.2 Interface shear and normal stresses

We recall that the finite element shear and normal stresses considered herein stem from the derivation of the finite element membranar and shear forces (finite element generalised forces).

The form of the shear stress, as determined analytically by the M4_5n, is well reproduced, for both of the two boundary conditions, except at the vicinity of the edge (see Figures 10). As regards M4_2n+1, the shear stress distributions are not as accurate as those provided by the M4_5n. Nevertheless, they do remain quite satisfactory, particularly with respect to the maximum values at the edges. The comparisons with Pagano's results also take place below.

We have also plotted the distributions of the $\nu^{0,90}$ and $\nu^{90,90}$ normal stresses, as given by the M4_5n, respectively at the 0/90 and 90/90 interface (see Figures 11). The high level of correlation between the curves is well established except, perhaps, at the 0/90 interface where a singularity has been noticed. In the case of micro-cracking, the finite element normal stresses also indicate singularities at the edge.

For the free edge problem, we have plotted the results given by Pagano's model which exactly verify the edge conditions; consequently, the shear stress is equal to zero on $y=b$. The maximum shear stress value occurs inside the ply and is less than that obtained by finite elements. Concerning the normal stress, the correlation between M4_5n (no normal stress in M4_2n+1) and Pagano's results is better, and this is for both interfaces. It is a very important point that our simplified models agree quite closely (for 2n+1) or even better (for 5n) with finite elements (in the confidence zone) than a more sophisticated model.

Moreover our models provide a very useful finite value on the edge, value to be improved by a delamination criterion (see Caron *et al.* (1999)). These two points reveal how well designed these simplified models are in studying edge effects.

In Figure 12, we have added results for a (±45°)$_s$, stemming from 3D calculations (that have not been described in this paper). The conclusions are similar and justify our choice of more straightforward 2D studies.

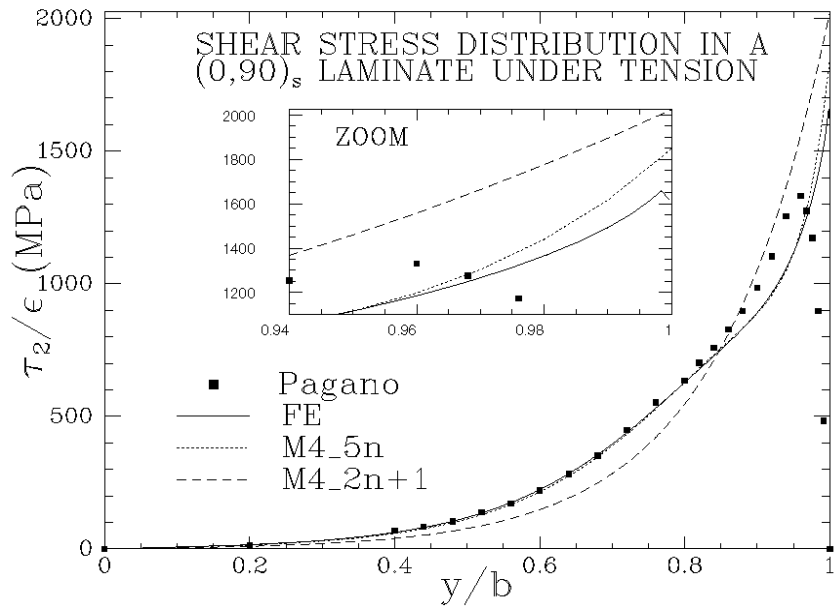

*Fig 10a : Shear stresses at the 0/90 interface. Free edge problem*

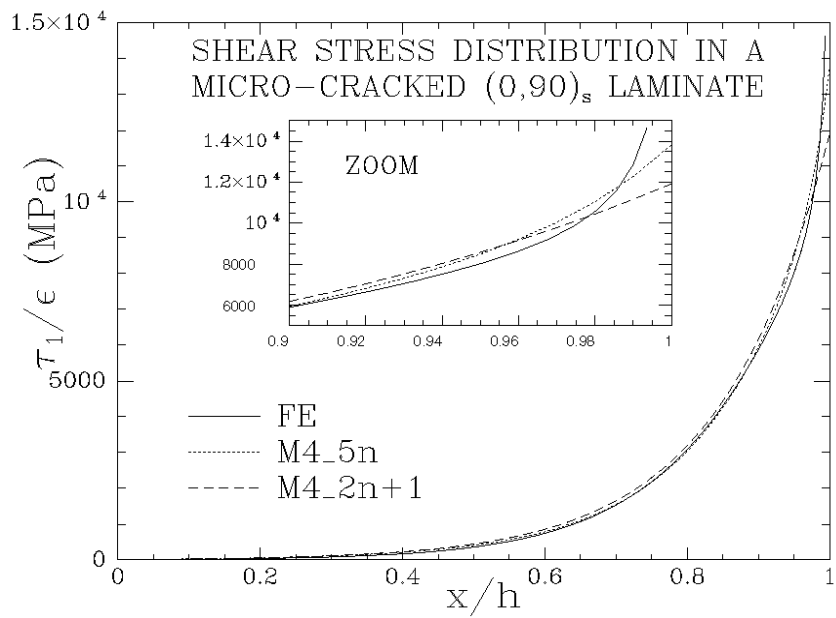

*Fig 10b : Shear stresses at the 0/90 interface. Micro-cracked cell*

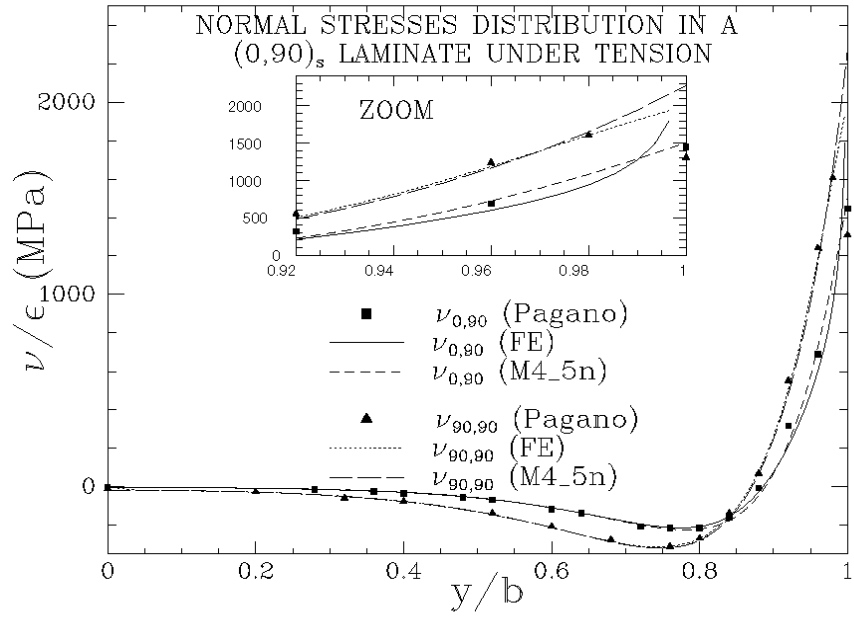

*Fig 11a : Normal stresses at the 0/90 and 90/90 interfaces. Free edge problem*

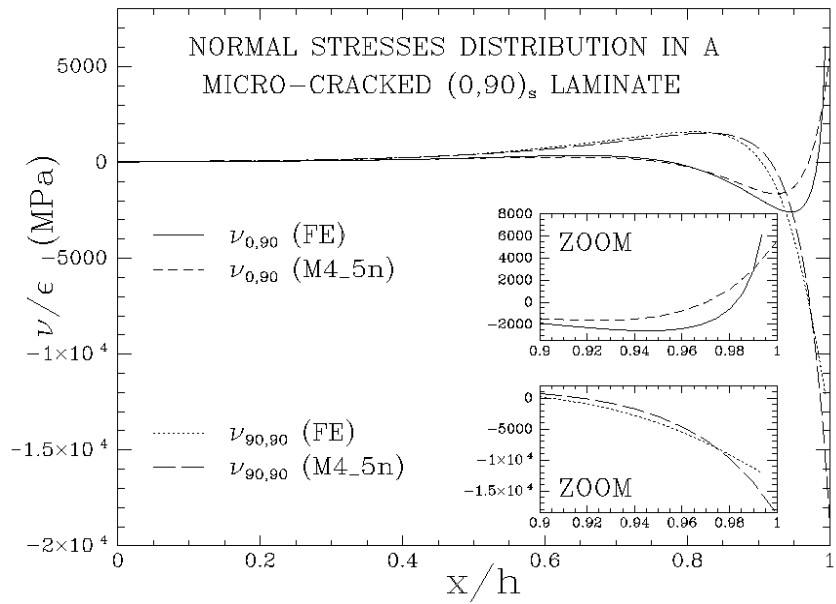

*Fig 11b : Normal stresses at the 0/90 and 90/90 interfaces. Micro-cracked cell*

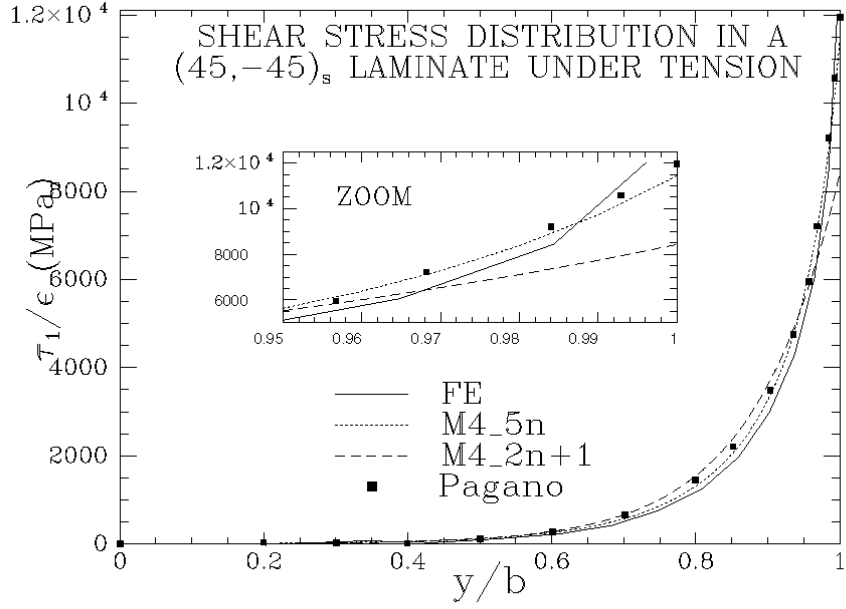

*Fig 12 : Shear stresses at the +45/-45 interface in a (±45)ₛ. Free edge problem*

## 6.3 Finite element and M4 energy comparison

It is now important to evaluate now our models through energy-related considerations.
Let's consider the 3D elastic energy associated with 3D stress fields.

$$W_{3D} = \sum_{i=1}^{n} \int_{h_i^-}^{h_i^+} \left[ \int_\omega \left( \frac{1}{2} \sigma_{mn} S_{mnop}^i \sigma_{op} \right) d\omega \right] dz \qquad (16)$$

where $S_{mnop}^i$ are the components of the ply $i$ compliance tensor.

### *M4_5n and finite element energy comparison :*

Our primary purpose here is to approximate the 3D energy using M4_5n generalised forces for the free-edge boundary conditions example. The elastic energy $W^{5n}$, associated with the approximated stress fields, can be written in the $y$-$z$ plane (we need only treat the quarter part of the laminate section and distinguish between the two plies) as follows:

$$W^{5n} = \int_{y=0}^{b} \left( w_M^{5n^{90}} + w_{M,\nu}^{5n^{90}} + w_\nu^{5n^{90}} + w_\tau^{5n^{90}} \right) dy + \int_{y=0}^{b} \left( w_M^{5n^0} + w_{M,\nu}^{5n^0} + w_\nu^{5n^0} + w_\tau^{5n^0} \right) dy \qquad (17)$$

where appearing in this order, the elastic energy due to membranar stresses, the membranar-normal coupling energy, the normal stress energy, and the shear stress energy. The expression of the shear stress energy is given as an example:

$$w_\tau^{5n\,i} = 4.\frac{1}{2}\left[\begin{array}{l}\dfrac{1}{e^i}Q_2^i S_{2323}^i Q_2^i + \dfrac{e^i}{12}\left(\tau_2^{i,i+1}-\tau_2^{i-1,i}\right)S_{2323}^i\left(\tau_2^{i,i+1}-\tau_2^{i-1,i}\right)\\[3mm]+\dfrac{1}{5e^i}\left(Q_2^i-\dfrac{e^i}{2}\left(\tau_2^{i,i+1}+\tau_2^{i-1,i}\right)\right)S_{2323}^i\left(Q_2^i-\dfrac{e^i}{2}\left(\tau_2^{i,i+1}+\tau_2^{i-1,i}\right)\right)\end{array}\right] \quad (18)$$

At this point, we can define the corresponding finite element energy $W^{EF}$, where $\sigma_{ij}$ are the finite element stresses values and $A^{el}$ the surface of an element $el$.

$$W^{EF}=\sum_{el\,\in\,90^\circ}A^{el}\left[w_M^{EF\,90}+w_{M,\nu}^{EF\ 90}+w_\nu^{EF\,90}+w_\tau^{EF\,90}\right]+\sum_{el\,\in\,0^\circ}A^{el}\left[w_M^{EF\,0}+w_{M,\nu}^{EF\ 0}+w_\nu^{EF\,0}+w_\tau^{EF\,0}\right] \quad (19)$$

where for each ply, we have introduced, membranar stresses energy, coupling energy, normal stress energy and shear stresses energy which is for example: $w_\tau^{EF\,0}=2S_{2323}^i\left(\sigma_{23}^i\right)^2$.

Table 2 compares the various energy contributions for the free edge problem in $(0,90)_s$ laminates. Differences between the two plies have also been studied :

| energy | M4_5n model (J) | 0° ply (%) | 90° ply (%) | F.E. model (J) | 0° ply (%) | 90° ply (%) |
|--------|-----------------|------------|-------------|----------------|------------|-------------|
| total | 7.55E-01 (100%) | 85.7 | 14.3 | 7.52E-01 (100%) | 85.1 | 14.9 |
| membranar | 6.87E-01 (91.1%) | 89.8 | 10.2 | 6.82E-01 (90.8%) | 89.6 | 10.4 |
| coupling | 0 (0%) | 0 | 0 | 1.36E-03 (0.2%) | 107.7 | -7.7 |
| normal | 1.92E-02 (2.5%) | 18.1 | 81.9 | 2.26E-02 (3%) | 8.7 | 91.3 |
| shear | 4.86E-02 (6.4%) | 55 | 45 | 4.52E-02 (6%) | 55.5 | 44.5 |

*Table 2 : free edge energy contributions in a $(0,90)_s$ laminate under tension*

The following essential conclusions can be drawn:

- It seems justifiable to consider the coupling energy negligible (it's actually an assumption of this model, which permits to obtain behaviour expressions).

- The shear energy distribution, which differs from the 0° ply to the 90° one, is accurately reproduced by the M4_5n model.

- The normal stress contribution is quite different for the two plies and this has basically been proved by the FE model. Nevertheless, if we now consider what we call an *interface energy (I.E.)* by summing normal and shear energies, we can note a very strong correlation between the two approaches (under brackets are the values for shear and normal finite element energies, respectively) :

$$I.E._{M4\_5n} = 6.78\text{E-}02\ J\ (= 1.92\text{E-}02 + 4.86\text{E-}02)$$

$$I.E._{FE}\ = 6.78\text{E-}02\ J\ (=2.26\text{E-}02 + 4.52\text{E-}02)$$

And for the microcracked cell,

$$I.E._{M4\_5n} = 7.84E\text{-}02\ J$$

$$I.E._{FE} = 7.42E\text{-}02\ J$$

### M4_2n+1 and finite element energy comparison

Our purpose now is to approximate the 3D energy using the M4_2n+1 generalised internal forces. Let us write the elastic energy $W^{(2n+1)M}$ associated to the M4_2n+1 approximated stress fields.

$$W^{2n+1} = \sum_{i=1}^{n} \iint_{\omega} \left( w_M^{(2n+1)^i} + w_\tau^{(2n+1)^i} \right) d\omega \tag{20}$$

The only difference with $5n$ energy concerns the expression of the shear stress energy :

$$w_\tau^{2n+1^i} = 4 \cdot \frac{1}{2} \left[ \frac{e^i}{12} \left( \tau_\alpha^{i,i+1} - \tau_\alpha^{i-1,i} \right) S_{\alpha3\beta3}^i \left( \tau_\beta^{i,i+1} - \tau_\beta^{i-1,i} \right) + \frac{e^i}{4} \left( \tau_\alpha^{i,i+1} + \tau_\alpha^{i-1,i} \right) S_{\alpha3\beta3}^i \left( \tau_\beta^{i,i+1} + \tau_\beta^{i-1,i} \right) \right] \tag{21}$$

We should emphasise that the coupling energy is still assumed to be negligible and that the normal interface energy is not present in this model (as a consequence of not introducing the corresponding generalised force).

If we compare the M4_2n+1 shear stress energy and the finite element interface energy for the free edge problem :

$$w_\tau^{2n+1} = 7.12E\text{-}02\ J \qquad vs \qquad I.E._{FE} = 6.78E\text{-}02\ J$$

and for the microcracked laminate :

$$w_\tau^{2n+1} = 7.62E\text{-}02\ J \qquad vs \qquad I.E._{FE} = 7.42E\text{-}02\ J$$

we would like to highlight once again the value of this *I.E.* concept, even for this simple model.

The energetic analysis we have performed in this section provides a better understanding of our model descriptions as well as an explanation of the meaning of a simplified model (e.g. just as the Love-Kirchhoff plate model is a simplified Reissner plate model, we can consider M4_2n+1 as a simplified M4_5n) : when a generalised force vanishes in a simplified model, this means that the associated energy is simply transferred or distributed into the other terms . In this way, a concept of interface energy, as the sum of energies due to shear and normal stresses, has been defined and validated.

## 7. Conclusion

This paper deals with the validation of simplified models which involve one kinematic per layer. We developed such a model that we named Multiparticle Models of Multilayered Materials (M4). These models allow introducing out-of-plane stresses (i.e. shear and normal stresses) at the interfaces of a laminate. We have examined in depth two of these models : the M4_$5n$ and the M4_$2n$+1. The first one can be described as a stacking sequence of Reissner-Mindlin plates, the second as a membranar superposition.

The validations are lead by means of finite element calculations in a $(0,90)_s$ laminate submitted to tension. Two boundary conditions were considered : free edge and microcracking. We also compare with results stemming from Pagano's works.

First of all, the validation procedure encountered numerical difficulties, since the finite element stresses are mean element values and not calculated at the right interface. For this reason, we introduced what we called finite element generalised forces, which are deduced from the 3D equilibrium equations and actually represent interface stresses. Next, a convergence study was conducted using these generalised forces, which are more relevant and stable tools. We could then focus on the singular behaviour of the stresses when approaching the edge or the microcrack; the maximum value of stresses thus depends upon the level of mesh refinement. Considering the nature of laminate edges (pulled out fibres, defects due to the elaboration process), we have proposed to define a zone of confidence (excluding a region where the calculated stresses have no meaning in relation to material heterogeneity) over which the convergence of the finite element generalised forces is ensured.

In comparing M4_$5n$ and finite element results in this zone of confidence, the following conclusions could be established : 3D stress fields are accurately reproduced even with critical boundary conditions (*i.e.* in the vicinity of a micro-crack) and the energy contributions in each ply, associated to the different stresses are calculated extremely well.

Validation was also carried out with the M4_$2n$+1 : due to the lower degree of the polynomials approximating the stress fields, this model is obviously less precise in describing stress distributions. The conclusion is that our simplified models seem to be very attractive, because more simple and more convenient for the study of edge effects, than more sophisticated ones.

With respect to the various energy contributions, our work has led to defining an *interface energy* by summing the two energies related to shear and normal stresses, respectively. A comparison with the same 3D energy proves to be very close: this point emphasises the fact that energetic approaches constitute a promising way to propose delamination criteria even with a simple model such as M4_2$n$+1.

Once again, we would like to insist that in the present work, all the conclusions and comparisons drawn between FE and analytical solutions concern the zone of confidence of laminates : nonetheless, the pertinence of the finite values on the edges of M4 interface forces shall be improved by an experimental approach (Caron *et al.* (1999)).